\documentclass{article}
\input amssym.def
\input amssym.tex
\def\ben{\begin{equation}}
\def\een{\end{equation}}
\def\half{{1 \over 2}}
\def\bea{\begin{eqnarray}}
\def\eea{\end{eqnarray}}
\def\p{\partial}

\def\bE{{\bf E}}

\def\babla{{\mbox{\boldmath $ \nabla $ }}} 
 \usepackage{epsfig}
\begin{document}

\hfuzz=100pt
\title{Spacetime interpretation of  
Torsion in Prismatic Bodies}
\author{G. Domokos \\
{\em Dept. Mechanics, Materials, and Structures,}\\
{\em Budapest University of Technology and Economics}\\
{\em M\"uegyetem rkp 3, H-1111 Budapest, Hungary}\\
{\em and}  \\
G. W. Gibbons  \\
{\em D.A.M.T.P., Cambridge University}\\
{\em Wilberforce Road, Cambridge CB3 0WA, U.K.}\\}
    
\maketitle

\begin{abstract}
A non-linear theory  for the plastic deformation
of prismatic bodies is constructed which 
interpolates between Prandtl's linear soap-film  approximation 
and N\'adai's  sand-pile model . Geometrically 
Prandtl's soap film and N\'adai's wavefront are unified into a 
single smooth surface
of constant mean curvature in three-dimensional Minkowski spacetime.

\end{abstract}

\section{Introduction} 

Prandtl's soap film \cite{Prandtl} and Nadai's \cite{Nadai0} 
sand heap analogies  remain
important tools for analysing the torsional loads
and plastic deformations 
of cylindrical shafts \cite{Chakrabarty,Alouges}. In this paper we propose
a generalisation of Prandtl's model in which the  eponymous potential  
$\phi(x,y)$ he introduced no-longer satisfies the linear Laplace equation
but rather   a non-linear equation known as the Born-Infeld equation.
This has the property that the  Prandtl potential $\phi(x,y)$ 
is continuous but nevertheless  the torsional  stress $|\nabla \phi|$ 
never exceeds
the plastic limit. The model
also permits a remarkable unified continuous  
geometrical analogue combining  both
the soap-film and sand pile analogies as limiting cases. 
According to this unified  analogy $\phi(x,y)$ defines a spacelike 
maximal surface in an auxiliary 2+1 dimensional spacetime.   
\section{The Prandtl-Nadai Construction}
In what follows we provide, for completeness,
a brief r\'esum\'e  of Prandtl's soap-film
and Nadia's sand heap analogy 
In the absence  of body forces, the governing equations are   
\ben  
\p_i T_{ij}=0\,\label{stress}
\een
 where $T_{ij}=T_{ji}$ is the {\it stress tensor}
(often often denoted  in Engineering texts $ \sigma_{ij}$ or $\tau_{ij}$
 cf. \cite{Chakrabarty}).
The {\it strain tensor} $e_{ij}$ is defined in terms of displacements
$u_i$ as $e_{ij}= \p_i u_j + \p _j u_i.$
 In linear theory for an isotropic substance Hooke's Law becomes
\ben
e_{ij}= {2 (1+ \nu )\over E} T_{ij} -{2 \nu \over E} \delta_{ij} T_{kk} \,.  
\een   
where $\nu$ is {\it Poisson's ratio}  and $E$ is {\it Young's modulus}. 
Following Saint-Venant\cite{StVenant} we assume that the  material is  
isotropic and  confined 
to a prismatic cylinder with base $B$ whose generators are 
parallel to the $z$  axis and orthogonal to the plane curve
$\gamma =\p B$ .
The displacements $(u,v,w)$  are assumed to satisfy  
$ u=-\theta y z\,
v=+\theta x z\,
w=w(x,y),$
where $\theta$ is a constant.
The function $w(x,y)$ is called the {\it warp function} 
and it gives the displacement parallel to the axis. 
These assumptions  imply
\ben
e_{xx}= 
e_{yy}=
e_{zz}=
e_{xy}=0, \quad
e_{yz}= \p_y w + \theta x, \quad
e_{zx}= \p_x w -\theta y. 
\een
Hooke's law implies that the diagonal components of the  stress tensor
(the {\it elongations})  vanish
and therefore  equation (\ref{stress})  gives
\ben
\p_x T_{xz} +\p_x T_{yz} =0 
\een
(this only requires $ \p_z T_{zz}=0$ ). We can 
now introduce the {\it Prandtl potential}  
\ben
T_{xz}= \p_y \phi  \,,\qquad T_{yz} =-\p_x \phi  \,. \label{pot}
\een
We have from (\ref{pot}) $\p_i \phi \, T_{i z} =0\,,i=x,y$
which implies that  the component of the force
in the $z$ direction   on the level sets of the 
Prandtl potential, whose normal is $\frac{\p_i \phi}{|\p _i \phi|}$,   vanishes.
Thus  any level set may act  as a  free boundary. It is convenient to set
$\phi=0$ on the boundary $\gamma=\p B$. 
The total torque is given by
\ben
T= \int_B  \bigl( x T_{yz} - y T_{xz}  \bigr) \, dx dy  = 2 \int _B \phi dx dy 
\een
on the use of the divergence theorem and the boundary conditions. If 
$G=E/2(1+\nu),$ and we use  Hooke's law we get 
\ben
T_{yz}= G(\theta y + \p_y w) \,, \qquad T_{xz}= G(-\theta x +\p_x w )\,.
\een
Thus the Prandtl potential satisfies Poisson's equation
\ben
{\p ^2\phi \over \p x^2 }  +  {\p ^2\phi \over \p y^2 }        =-2G \theta.
\label{Poisson} \een
This must be solved in $B$, subject to the boundary condition that
$\phi=0$ on the boundary $\gamma =\p B$.
Of course Poisson's equation 
arises in electrostatics where the electric field, and electrostatic
potential $\phi$ satisfy  
 $ \bE = -\babla \phi $ ,
$ \babla \cdot  \bE = \rho$, where $\rho$ is the density of electric charge.
In  the case of Prandtl's equation (\ref{Poisson}) 
we have $\rho=2 G \theta$ which corresponds to constant charge density.     
Prandtl adopted a different analogy.
He  regarded  $\phi =\phi(x,y) $ as a height function
in some auxiliary three dimensional space Euclidean space 
$\Bbb E ^3 $  with coordinates $(x_1,x_2,x_3)= (x,y,\phi)$ 
then the surface is {\sl approximately}  of constant  mean curvature
such as would be adopted by a soap-film
with a constant pressure difference.
The torque is then proportional to  the volume
between the approximate soap film and the base $B$.
Note that if the analogy were {\sl exact} then  the Prandtl potential would
satisfy the Young-Laplace equation
\ben
\p_x \frac{\p_x \phi }{\sqrt{ 1 +|\nabla \phi|^2 }} 
+ \p_y \frac{\p_y\phi }{\sqrt{ 1 + |\nabla \phi|^2 }} =-2 G \theta \,. \label{soap}
\een

\subsection{Non-linear theory}

If the total shear stress exceeds the elastic
limit $k$, then the material deforms plasticly.
Thus a {\it plastic upper  bound}  must be satisfied
\ben
T_{xz}^2 + T_{yz} ^2 = | \nabla \phi |^2  \le k^2\,, \label{bound}
\een
where $k$ is the plastic limit
and in the plastic region the Prandtl potential is assumed to  satisfy the {\it Eikonal equation}     
\ben
| \nabla \phi |^2  = k^2 \,. \label{Eikonal} 
\een
According to {\it N\'adai's sandpile analogy} \cite{Nadai0,Nadai1,Nadai2}
one may regard  a  Prandtl potential satisfying the Eikonal equation
(\ref{Eikonal})  as the height function $\phi$
of a sand-heap or sand-pile of constant {\it angle of repose}  
$\alpha = \tan^{-1} k\,$   
located in same Euclidean space ${\Bbb E} ^3$. 
The strategy for solving the for $\phi$ adopted by Prandtl
is to erect over the base $B$, a sand-pile, that is the  solution 
$\phi_{\rm Eikonal} $
of the Eikonal equation (\ref{Eikonal}). One then also erects the approximate
soap film, that is the graph
of  the 
solution of Poisson's equation (\ref{Poisson}) $\phi_{\rm Poisson} $. 
The solution adopted by Prandtl is 
then to take the \em minimum \rm of $\phi_{Poisson}$ and  $\phi_{Eikonal}.$ 
In other words, linear theory is assumed valid under the tent.

\section{Spacetime Interpretation} 

We begin by noting that the the Eikonal equation (\ref{Eikonal}) 
may be regarded as the defining a {\it wave-front, null
or characteristic surface} ${\cal N} : u(x,y,t) =0 $ where 
\ben
u=t- \phi(x,y) =0  \label{surface}
\een 
in {\it $3$-dimensional Minkowski spacetime} ${\Bbb E} ^{2,1}$ 
with coordinates $x,y,t$ and metric   
\ben \label{Minkowski2}
ds ^2 = dx^2 + dy ^2 - c^2 dt ^2,
\een
where $c= 1/k$ is the velocity of light. In other words,
\ben
-\frac{1}{c^2} (\p_t u)^2 +(\p_x u)^2 + (\p_y u) ^2 =0\,.
\een
In general, the plastic upper bound (\ref{bound}) must hold. 
However the transition between linear and plastic behaviour
is observed to be smoother than the abrupt change envisaged in  
Prandtl's theory.   
Now (\ref{bound})   implies that the surface $\Sigma \subset {\Bbb E} ^{2,1}$
has a normal which is non-spacelike. If inequality holds
in (\ref{bound}) then  $\Sigma$ given by (\ref{surface}) 
is a {\it spacelike surface}. Thus one considers a smooth surface
whose normal may become null, i.e., may satisfy (\ref{Eikonal}) in a continuous fashion.        
It is natural therefore to replace the linear Poisson
equation with a non-linear equation which interpolates
between the Poisson equation and the Eikonal equation
and which has  an interpretation more in keeping
with the Minkowski spacetime framework. On might try
replacing Poisson's equation by the Young-Laplace equation (\ref{soap}) 
but this does not fit well with our interpretation the Eikonal equation.
A geometrically better motivated  suggestion 
is to  postulate  the relativistically covariant analogue of the
Young-Laplace  equation: 
\ben
\p_x \frac{\p_x \phi }{\sqrt{ 1- c^2 |\nabla \phi|^2 }} 
+ \p_y \frac{\p_y\phi }{\sqrt{ 1- c^2 |\nabla \phi|^2 }} =-2 G \theta \,, 
\label{mean}
\een  
which arises as the Euler- Lagrange equations
of the functional 
\ben
\int _B \Bigl( \sqrt{ 1- c^2 |\nabla \phi|^2 } +2 G c^2  \theta  \phi \Bigr ) \,dx dy \,.  
\label{action}
\een  
The first term in (\ref{action}) is the area of any spacelike surface
$\Sigma$  
with edge  $\gamma = \p B$ and the second the spacetime volume
between $\Sigma $ and the spacelike hyperplane of constant time $t=0$.    
It follows that (\ref{mean}) describes a {\it spacelike surface of constant 
mean curvature in three dimensional Minkowski spacetime ${\Bbb E} ^{2,1}$}. 
The  difference in sign in the arguments inside the  square roots
in the denominators of  (\ref{soap}) and (\ref{mean}) is 
intended to enforce the plastic bound (\ref{bound}). 
\section{Born and Infeld's non-linear  Electrodynamics}
Just as Poisson's equation (\ref{Poisson})  has an electrostatic analogue,
so does our proposed replacement (\ref{mean}). 
The equation (\ref{mean}) and associated variational principle
with functional  (\ref{action}) admit a similar   physical interpretation
\cite{Gibbons}.
Born and later Born and Infeld \cite{BornInfeld,Born}
suggested a non-linear version of 
electrodynamics: a modification of Maxwell's equations
in which there is a maximum electric field strength.
In the electrostatic situation the electric field strength ${\bf E}$ 
electric induction ${\bf D}$ and charge density $\rho$ 
satisfy ${\bf E}= - \nabla \phi\,{\nabla} \cdot {\bf D} = \rho \,  
{\bf D}={\bf E}/(\sqrt{1-{\bf E}^2/b^2 }).$   
In two dimensions, these equation, with $\rho= 2G \theta$, $b=1/c$,
coincide with (\ref{mean}) 
and were first studied in the context of Born-Infeld
electrostatics by Pryce \cite{Pryce1,Pryce2}.

\section{Comparison in the circular case}

\begin{figure}
\begin{center}
\includegraphics[width=110 mm]{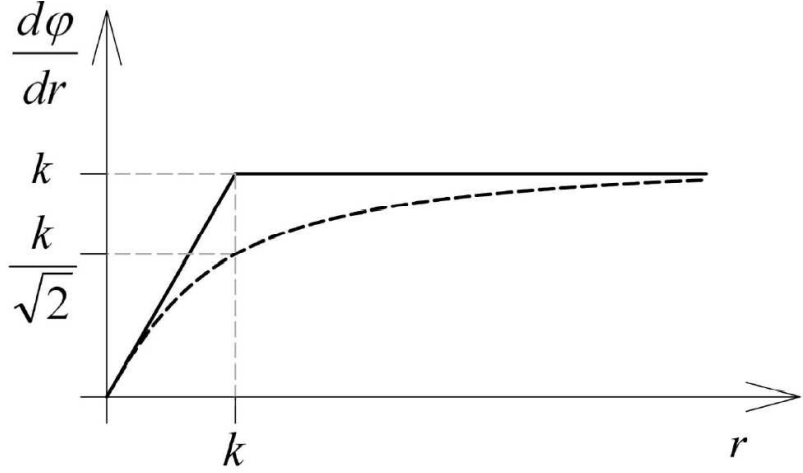}
\end{center}
\caption{Comparison between the two models. Solid line: Prandtl-Nadai theory, dashed line $d\phi/dr= G\theta rk/\sqrt{k^2+G^2\theta^2r^2}$: our model. Maximal deviation at
elasto-plastic boundary. The dashed curve also yields the constitutive law if we subsitute $\gamma=r\theta$ for the strains.}\label{Fig1}
\end{figure}
If the cross section is circular, the standard case has
\bea
\phi&=& -\frac{G\theta}{2} r^2 + \phi_0\,,  \qquad  0\le r \le  \frac{k}{G \theta}\\
&=& -kr + \frac{k^2}{2G \theta} +\phi_0 \,,\qquad r> \frac{k}{G\theta} \,.
\eea
The inner region is the Prandtl regime. The outer region is
the N\'adai's sand pile  regime. The constant $\phi_0$ is chosen
so that $\phi(r=R)=0$, where $R$ is the radius of the cylinder. 
Thus if $R< \frac{k}{G\theta} $, then $\phi_0= \frac{G \theta}{2} R^2 $
and Prandtl's solution is $\phi=  -\half G \theta (  r^2- R^2)$  
while if $R >  \frac{k}{G\theta} $, then $\phi_0= kR - \frac{k^2}{2G \theta} $
and the sandpile solution is conical with $\phi= -k(r-R) $ . 
In the Born-Infeld case one has a single unified formula

\ben
\phi= \frac{k^2}{G \theta} \Bigl ( \sqrt{ 1 + \frac{G^2 \theta ^2 R^2}{k^2} 
}   -\sqrt{  1 + \frac{G^2 \theta ^2 r^2}{k^2} 
}   \Bigr ) \,.
\een
For small $r$, we expand both square roots and recover
Prandtl's solution at lowest order.  For large $R$ and hence 
large $r$ we ignore the one
inside the square roots and the solution  approachs the conical
sand pile solution. Note that while $\phi(R)=0$ 
in both solutions,  in general  the values of $\phi(0)$
are not the same. The predictions of both theories are compared in Figure \ref{Fig1}.
The total torque $T=4\pi\int_0^R r\phi\, dr$ is given by
\ben
T= 4 \pi \frac{k^2}{G \theta} 
 \Bigl (\half R^2 \bigl (  1 + \frac{G^2 \theta ^2 R^2}{k^2} \bigr )^\half
+ \frac{k^2}{3 G^2 \theta ^2} \bigl ( 1-  
(1+ \frac{G^2 \theta ^2}{k^2} R^2 ) ^ {\frac{3}{2} }  \bigr )      \Bigr). 
\een
\section{Interpretation in Minkowski spacetime}

We have 
\ben
\bigl ( \frac{\phi-\phi_0}{k} \bigr ) ^2  - x ^2 - y^2  = \frac{k^2}
{G^2 \theta ^2} \,. 
\een 
Thus if we think of $( \phi, x,y ) $ as coordinates for three dimensional
$(x,y,\phi)$ Minkowski spacetime with metric given in (\ref{Minkowski2}),
then $\phi$ should be thought of as the time coordinate.
Our solutions is now seen as a hyperboloid of constant spacetime distance from
the point  $(\phi_0, 0,0 ) $, i.e the analogue in Minkowski
space of a sphere in Euclidean space. To get from the sphere
to the hyperbola one may ``Wick Rotate'', i.e., set
\ben
c (\phi - \phi_0)  = i (z-z_0)G \theta/k = i a^{-1}, 
\een
so that  $(z-z_0) ^2 + x^2 + y^2 = a ^2.$
The fact that the mean curvature (in the Lorentzian sense) is
constant is now obvious because the surface 
is invariant under Lorentz transformations
about $(\phi_0, 0,0 ) $ and the mean curvature is a Lorentz scalar.

\section{Acknowledgements}
The first author was a Visiting Fellow Commoner at Trinity College, Cambridge
during part of this collaboration.
This research was supported by the Hungarian National Foundation (OTKA) Grant K104601.
The authors thank Tam\'as Ther for his help with Figure 1.

\end{document}